\documentclass[aps,prl,twocolumn,superscriptaddress]{revtex4-2}
\usepackage{bbm}
\usepackage{mathrsfs}
\usepackage{amsmath}
\usepackage{amsfonts}
\usepackage[colorlinks=true,linkcolor=blue,urlcolor=blue,citecolor=blue,anchorcolor=blue]{hyperref}
\usepackage{graphicx,epstopdf}
\usepackage{subfigure}
\usepackage{epsfig}
\usepackage{dcolumn}
\usepackage{bm}
\usepackage{color}
\usepackage{natbib}
\usepackage{amssymb}
\usepackage{xcolor}
\usepackage{braket}
\usepackage{ulem}
\usepackage{float}
\usepackage{lipsum}
\usepackage{array}
\usepackage{booktabs}

\begin{document}

\title{Heisenberg-Scaling Quantum Metrology of Electron-Photon Coupling via Multi-Electron PINEM}
\author{Zihang Zou}
\address{State Key Laboratory of Artificial Microstructure and Mesoscopic Physics, School of Physics, Frontiers Science Center for Nano-optoelectronics, $\&$ Collaborative Innovation Center of Quantum Matter, Peking University, Beijing 100871, China}
\author{Feng-Xiao Sun}
\email{sunfengxiao@bupt.edu.cn}
\address{State Key Laboratory of Information Photonics and Optical Communications, Beijing Key Laboratory of Information Metamaterials, $\&$ School of Physical Science and Technology, Beijing University of Posts and Telecommunications, Beijing 100876, China}
\address{State Key Laboratory of Artificial Microstructure and Mesoscopic Physics, School of Physics, Frontiers Science Center for Nano-optoelectronics, $\&$ Collaborative Innovation Center of Quantum Matter, Peking University, Beijing 100871, China}
\author{Yunquan Liu}
\address{State Key Laboratory of Artificial Microstructure and Mesoscopic Physics, School of Physics, Frontiers Science Center for Nano-optoelectronics, $\&$ Collaborative Innovation Center of Quantum Matter, Peking University, Beijing 100871, China}
\address{Collaborative Innovation Center of Extreme Optics, Shanxi University, Taiyuan, Shanxi 030006, China}
\author{Qiongyi He}
\email{qiongyihe@pku.edu.cn}
\address{State Key Laboratory of Artificial Microstructure and Mesoscopic Physics, School of Physics, Frontiers Science Center for Nano-optoelectronics, $\&$ Collaborative Innovation Center of Quantum Matter, Peking University, Beijing 100871, China}
\address{Collaborative Innovation Center of Extreme Optics, Shanxi University, Taiyuan, Shanxi 030006, China}
\address{Hefei National Laboratory, Hefei 230088, China}

\begin{abstract}
    Free-electron quantum optics has recently emerged as a promising platform for coherent electron-light interactions, yet its potential for quantum-enhanced metrology remains largely unexplored. Here, we investigate the fundamental precision limit of coupling-strength estimation in multi-electron photon-induced near-field electron microscopy (PINEM), demonstrating that the quantum Fisher information scales quadratically with electron number. We then propose an experimentally accessible scheme combining squeezed-light probing and optical homodyne detection that attains Heisenberg scaling with respect to electron number, provided the electrons exhibit nonzero inter-electron coherence, while surpassing conventional PINEM sensing in photon-number enhancement as well. We further analyze the effect of imperfect electron preparation, showing that electron phase noise is converted into a fluctuating optical displacement and eventually limits the precision. To overcome this limitation, an auxiliary optical mode coupled to the same electron ensemble measures the displacement noise in the signal mode, allowing its conditional suppression and recovery of the Heisenberg scaling. Our work establishes multi-electron coherence as a resource for free-electron quantum metrology and opens the route to precision measurements of nanoscale electron-light interactions.
    
\end{abstract}

\maketitle


Free-electron physics has developed rapidly in recent years, supporting significant advances in applications that range from particle accelerators~\cite{acce-1,acce-2,England2014Dielectric} and free-electron lasers~\cite{free-electron-laser-2,acce,free-electron-laser} to ultrafast electron microscopy~\cite{UEM1,UEM2,UEM3}. In particular, photon-induced near-field electron microscopy (PINEM)~\cite{PINEM-source-exp,PINEM-source,Garcia2010Multiphoton,PhysRevLett.122.183204} enables coherent interactions between free electrons and optical near fields~\cite{coupling-strong,coupling-strong1,coupling-strong2,coupling-strong3,Xie2025Maximal}, establishing free-electron quantum optics as a promising platform for generating and manipulating quantum states of light and matter~\cite{Free-electron-quantum-optics,shi_quantum_2024,de2025roadmap}. Within this framework, considerable effort has been devoted to electron-photon entanglement~\cite{ent-2019,Armin2022Cavity,baranes2022free,ele-pho-ent1}, nonclassical state preparation~\cite{create-cat-gkp,create-cat,QPINEM,Structured-electrons}, quantum state tomography~\cite{tomography,Imprinting-the-quantum-statistics-of-photons-on-free-electrons}, and free-electron qubits with ultrafast quantum gates~\cite{Reinhardt2021Qubits,free-ele-computing,ele-GKP,PhysRevLett.132.035001}. These advances make free electrons a versatile quantum resource that interfaces optical quantum states and nanoscopic systems on ultrafast timescales, naturally motivating their application to quantum metrology.

Quantum metrology exploits nonclassical resources to overcome the standard quantum limit of classical estimation strategies and approach the Heisenberg limit, the ultimate precision bound imposed by quantum mechanics~\cite{qmetro-rev,PARIS2009QUANTUM,giovannetti2011advances,Xiang2013}. In free-electron platforms, quantum sensing has so far been investigated to estimate the electron-photon coupling strength based on quantum Mach-Zehnder interferometry~\cite{free-ele-phase-modu,Chen2023Quantum,free-electron-noon}, as well as free-electron Ramsey-type interferometry that provides simultaneous access to the amplitudes and phases of sample responses~\cite{FERI,g_ref,bucher2024coherently}. However, existing PINEM-based estimation protocols predominantly employ single-electron probes and classical light, and their achievable precision is therefore fundamentally limited by the absence of both the collective coherence of multi-electron states and the quantum resources of nonclassical optical fields. Although theoretical frameworks have recently been established for correlated multi-electron states and their interactions with light~\cite{multielectron-model,multielectron-spin-squeezing,two-electron-quantum-walk}, it remains an open question how to combine collective electron coherence with quantum optical resources for quantum-enhanced measurement, and thereby approach the ultimate metrological limit of free-electron platforms. 

In this Letter, we prove that the quantum Fisher information (QFI) of multi-electron PINEM reaches the Heisenberg scaling with the electron number $F_Q\sim N_e^2$. To achieve this limit, we propose a homodyne-based scheme in which $N_e$ comb-state electrons interact via PINEM with a squeezed optical mode of average photon number $N_{ph}$, and the resulting optical field is measured by homodyne detection. For ideally prepared electron comb states with nonzero inter-electron coherence, the measurement sensitivity scales as $\chi^{-2}\propto N_e^2N_{ph}$, saturating the quantum Cram\'{e}r-Rao bound and providing an enhanced photon-number scaling compared to conventional schemes. However, imperfect electron-state preparation gives rise to classical phase fluctuations that degrade the achievable sensitivity. To tackle this challenge, we introduce an auxiliary optical mode and conditional measurements, where the ultimate measurement sensitivity can be recovered even for imperfect electron states. 

We first evaluate QFI $F_Q$ to determine the ultimate precision limit for estimating the electron-photon coupling strength $g_s$ in multi-electron PINEM. Assuming the coupling strength is encoded through a process with generator $\hat G$, where $\rho(g_s)=\exp(ig_s \hat G)\rho \exp(-ig_s\hat G)$, the estimation precision is limited by the quantum Cram\'{e}r-Rao bound, $\Delta^2 g_s\ge 1/F_Q$~\cite{CR-bound}. Thus, the measurement sensitivity defined as $\chi^{-2}=1/\Delta^2 g_s$~\cite{Fadel_2025} is bounded by the QFI. In particular, the QFI is reduced to $F_Q=4\Delta^2\hat G$ for pure probe states, where $\Delta^2\hat G=\langle \hat{G}^2 \rangle-\langle \hat{G} \rangle^2$ is the variance.

The multi-electron PINEM interaction is described by
\begin{equation}
U(g_s)=
\exp\!\left[
\sum_{i=1}^{N_e}
\left(
g_sa^\dagger b_i-g_s^*ab_i^\dagger
\right)
\right],
\label{eq:scattering}
\end{equation}
where $a$ is the optical annihilation operator and $b_i|E\rangle_i=|E-\hbar\omega\rangle_i$ is the energy-ladder operator of the $i$th electron. Decomposing $g_s=g_x+ig_y$, we treat its real and imaginary parts as independent parameters, with corresponding generators
$\hat G_x=-i\sum_i(a^\dagger b_i-ab_i^\dagger)$ and $\hat G_y=\sum_i(a^\dagger b_i+ab_i^\dagger)$, respectively. The multi-electron PINEM interaction can then be rewritten as $U(g_s)=\exp[i(g_x\hat G_x+g_y \hat G_y) ]$.
Hence, for $N_e$ independently prepared electrons in the same quantum state, the respective QFI takes the form of
\begin{equation}
F_Q^{(\mu)}
=
4N_e\Gamma_{ii}^{(\mu)}
+
4N_e(N_e-1)\Gamma_{ij}^{(\mu)},
\qquad
\mu=x,y,
\label{eq:qfi}
\end{equation}
where $\Gamma_{ij}^{(\mu)}=\langle d_{i}^{(\mu)}d_{j}^{(\mu)} \rangle-\langle d_{i}^{(\mu)} \rangle\langle d_{j}^{(\mu)} \rangle$ with $d_{i}^{(x)}=i(a^{\dagger}b_i-ab_i^{\dagger})$ and $d_{i}^{(y)}=a^{\dagger}b_i+ab_i^{\dagger}$. Accordingly, the inter-electron covariance $\Gamma_{ij}^{(\mu)}$ is responsible for the collective enhancement of the QFI and the necessary condition for nonzero $\Gamma_{ij}^{(\mu)}$ is $\langle b_i\rangle\ne 0$ for each electron. In addition, since $\Gamma_{ij}^{(\mu)}$ is in the order of $N_{ph}=\langle a^\dagger a\rangle$, the QFI reaches the scaling of $N_e^2 N_{ph}$ (see the Supplemental Material (SM)~\cite{supplementary} for details). 


The mechanism underlying this enhancement is readily illustrated when all electrons are prepared in the comb state $|{\rm comb}(\phi)\rangle=\sum_ne^{in\phi}|n\rangle$, which satisfies
$b_j|{\rm comb}(\phi)\rangle=e^{i\phi}|{\rm comb}(\phi)\rangle$.
For the initial state
$|\Psi_0\rangle=|{\rm comb}(0)\rangle^{\otimes N_e}\otimes|\psi_{\rm ph}\rangle$,
the PINEM interaction reduces to
\begin{equation}
U(g_s)|\Psi_0\rangle
=
|{\rm comb}(0)\rangle^{\otimes N_e}
\otimes
D(N_eg_s)|\psi_{\rm ph}\rangle,
\label{eq:displacement}
\end{equation}
where $D(\alpha)=\exp(\alpha a^\dagger-\alpha^*a)$ is the optical displacement operator with amplitude $\alpha$. Thus, multi-electron PINEM encodes $g_x$ and $g_y$ in the optical displacement of the amplitude $\alpha=N_eg_s$, leading to $F_Q(g_x) \propto N_e^2$ and $F_Q(g_y) \propto N_e^2$.
This result directly reveals Heisenberg scaling in multi-electron PINEM metrology. In the following, we show that this precision limit can be attained with a practical homodyne-detection protocol.


\begin{figure}
    \centering
    \hspace{0.0cm}
    \includegraphics[width=1.0\linewidth]{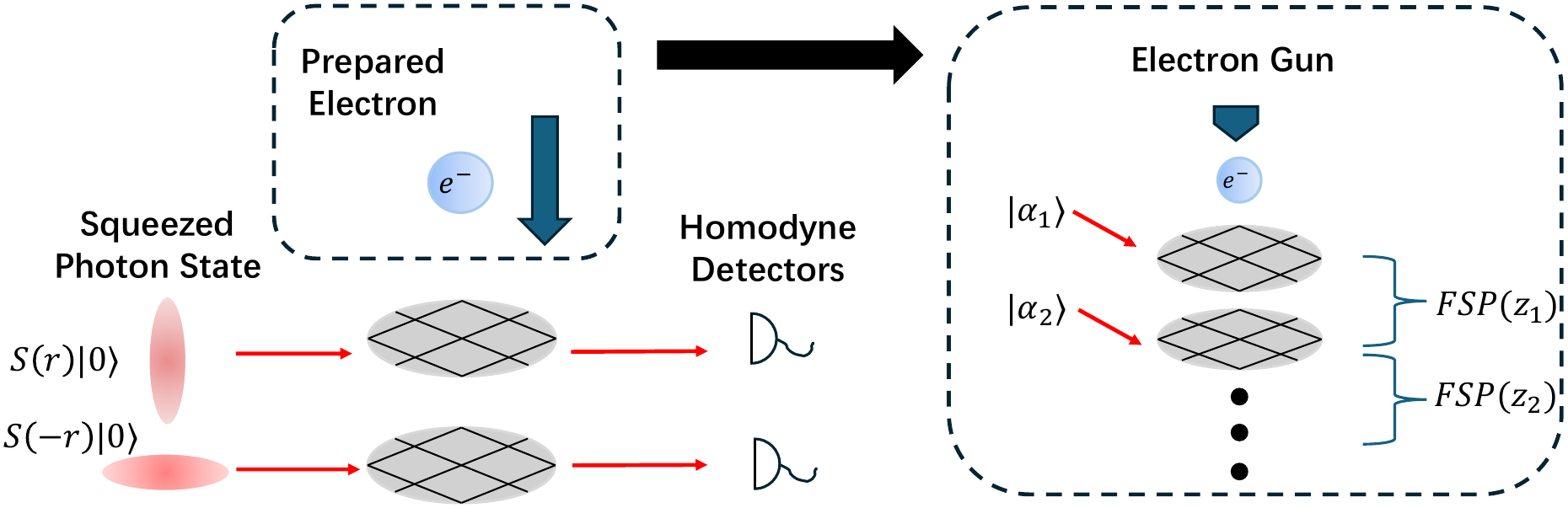}
    
    \caption{Experimental apparatus for multi-electron PINEM metrology. A train of $N_e$ electrons emitted from an electron gun is prepared in identical coherent comb states by cascaded interactions with classical light fields and free-space propagation (FSP) with the method developed in Ref.~\cite{coherent-electron}. The prepared electrons subsequently interact, through the PINEM coupling $g_s$, with two orthogonally squeezed states, $S(r)|0\rangle$ and $S(-r)|0\rangle$. Homodyne detection is applied to the output modes therefore provides independent estimates the real and imaginary parts of the coupling strength $g_s$.}
    \label{multi-electron-exp}
\end{figure}

The proposed measurement scheme is illustrated in Fig.~\ref{multi-electron-exp}, where a train of $N_e$ electrons is prepared in identical coherent comb states satisfying $\langle b_j\rangle\simeq1$, following the previously reported protocol composed of cascaded interactions with classical light and free space propagation (FSP)~\cite{coherent-electron}. The electrons subsequently interact with two squeezed vacuum states through PINEM coupling, and the electron-photon coupling strength is estimated through homodyne measurements of the output optical states.
To independently estimate the real and imaginary parts of $g_s$, the two optical modes are prepared in orthogonally squeezed vacuum states. The first mode is supposed to be an amplitude-squeezed state $S(r)|0\rangle$ where $S(r)=\exp[\frac r2 (a^2-a^{\dagger 2})]$. Thus, for its output state  $D(N_eg_s)S(r)|0\rangle$, homodyne detection of the amplitude-quadrature $X=a+a^\dagger$ provides an estimator for the real part $g_x$ with $\langle X\rangle=2N_e g_x$.
Similarly, for the second mode as a phase-squeezed state $S(-r)|0\rangle$, measurement of the phase-quadrature $P=i(a^\dagger-a)$ gives $\langle P\rangle=2N_e g_y$ for evaluating the imaginary part $g_y$.
Thus, the measurement sensitivity for $g_\mu(\mu=x,y)$ is determined by~\cite{Fadel_2025}
\begin{equation}
\chi^{-2}_\mu=\frac{|\partial_{g_\mu}\langle O_\mu\rangle|^2}{\Delta^2 O_\mu},
\end{equation}
with $O_x=X$ and $O_y=P$.

For equal squeezing parameters $r$ in the two optical modes, the sensitivities for estimating $g_x$ and $g_y$ are identical, $\chi^{-2}=\chi^{-2}_x=\chi^{-2}_y=4e^{2r}N_e^2$, where the respective variance satisfies $\Delta^2O_\mu=e^{-2r}$. In addition, the average photon number $N_{\rm ph}\propto e^{2r}$ for squeezed vacuum states in the strong squeezing regime, which is achievable due to recent experimental advances~\cite{Sh.Iskhakov:12,Perez:14,Lyu2026Generation}. Thus, the sensitivity results in $\chi^{-2}\propto N_e^2N_{\rm ph}$ matching the QFI scaling. Therefore, our experimentally accessible measurement scheme fully exploits the quantum enhancement arising from coherent multi-electron PINEM interactions.

\begin{figure}[t]
\centering
\includegraphics[width=1.0\linewidth]{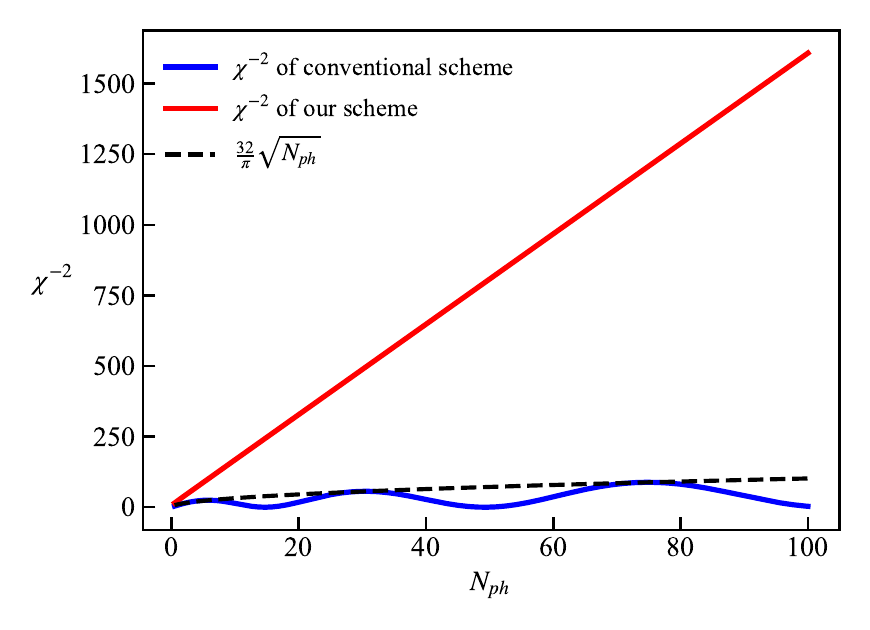}
\caption{Sensitivity scaling with photon number for the conventional and our schemes, compared in the single-electron regime with $g_s=0.5$. The sensitivity $\chi^{-2}$ of each scheme is plotted against the average photon number $N_{ph}$ for a conventional scheme determined by a coherent field of amplitude $\alpha$ ($N_{ph}=|\alpha|^2$, blue) and for our scheme $N_{ph}$ determined by squeezing parameter $r$ ($N_{ph}=\sinh^2 r$, red). The sensitivity of the conventional scheme is bounded by the envelope $\mathrm{Env}(\chi^{-2})=\frac{16}{\pi g_s}\sqrt{N_{ph}}=\frac{32}{\pi}\sqrt{N_{ph}}$ at $g_s=0.5$ (black dashed line). By contrast, combining electron-comb coherence with a squeezed optical probe yields a linear scaling $\chi^{-2}\propto N_{ph}$, so that our scheme already offers an advantage over the conventional scheme in the single-electron case.}
\label{single-electron}
\end{figure}


Now, we will benchmark the metrological potential of our proposal against that of standard PINEM. Noting that in conventional PINEM schemes only a single electron is considered, which is prepared in its energy eigenstate and then interacts with a coherent optical field of amplitude $\alpha$, we thus compare the resulting sensitivities in the single-electron regime. In conventional PINEM schemes, the probability of detecting the electron in the $n$th energy sideband is $P_n=|J_n(2|g\alpha|)|^2$, where $J_n$ denotes the Bessel function of the first kind. According to its measurement sensitivity  $\chi^{-2}=|\partial_gP_n|^2/[P_n(1-P_n)]$, the large-argument asymptotics of the Bessel function yield $\chi^{-2}\approx(16\alpha/\pi g)\sin^2(2g\alpha-n\pi/2-\pi/4)$. The oscillatory factor describes the sideband-dependent modulation, while the envelope scales as $\mathrm{Env}(\chi^{-2})\propto\alpha$. For coherent states $N_{\rm ph}=|\alpha|^2$, the conventional PINEM scheme therefore exhibits a sensitivity scaling of $\chi^{-2}_{\rm conv}\propto\sqrt{N_{\rm ph}}$ (See SM~\cite{supplementary} for details). 
Recall that in our scheme $\chi^{-2}_{\rm comb}\propto N_{\rm ph}$ with a single comb-state electron. Therefore, even without the multi-electron enhancement, combining electron-comb coherence with a nonclassical optical probe provides a superior scaling compared with conventional PINEM measurements, which is illustrated in Fig.~\ref{single-electron} summarizing the respective photon-number scalings of these two approaches.

We now return to the effects of multi-electron coherence. In conventional PINEM schemes where only a single electron is involved at a time, the electron number $N_e$ is exactly the number of experimental repetitions, so the QFI increases linearly with $N_e$. Hence, Our scheme offers an advantage in terms of the number of electrons as the Heisenberg scaling can be obtained with ideally prepared coherent electron comb states. However, in realistic experiments, imperfect electron preparation processes introduce phase uncertainty and reduce the coherence of the electron ensemble. We next analyze how such imperfections affect the achievable metrological sensitivity.

Since the comb state is unchanged under the PINEM interaction, an imperfectly prepared electron state can be modeled as a statistical mixture of comb states,
\begin{equation}
\rho_e
=
\sum_\phi p_\phi
|{\rm comb}(\phi)\rangle
\langle{\rm comb}(\phi)| ,\phi\in[-\pi,\pi],
\end{equation}
where $p_\phi$ denotes the phase distribution of the electron ensemble. After the PINEM interaction and tracing out the electrons, the effective optical state becomes a statistical mixture of the displaced squeezed states $D(\alpha)S(r)\ket 0$, with displacement amplitude $\alpha=g_s\sum_{i=1}^{N_e}e^{i\phi_i}.$ Therefore, the electron phase fluctuations are directly transferred into the optical displacement noises. To estimate $\text{Re}(g_s)$, we perform homodyne detection of the amplitude quadrature and define $\xi=2\mathrm{Re}(g_s\sum_i e^{i\phi_i})$, where the phases $\phi_i$ are independently sampled from $p_\phi$. The resulting quadrature expectation value $\langle X\rangle=\langle \xi\rangle$, while the variance contains two contributions, $\Delta^2X=e^{-2r}+\Delta^2\xi$. Here, $e^{-2r}$ corresponds to the quantum noise of squeezed light, while $\Delta^2\xi$ originates from statistical fluctuations in the electron preparation. When $\text{Im}(g_s)=0$, the corresponding sensitivity becomes
\begin{equation}
\chi^{-2}
=
\frac{
4N_e^2\langle\cos\phi\rangle^2
}{
e^{-2r} + 4N_e \Delta^2 \text{Re}(g_se^{i\phi})
}.
\label{eq:imperfect_sensitivity}
\end{equation}

Equation (\ref{eq:imperfect_sensitivity}) reveals two distinct regimes. When the electron number is relatively small, the squeezing noise dominates and the sensitivity preserves the ideal scaling, $\chi^{-2}\propto N_e^2N_{\rm ph}$.
However, for sufficiently large $N_e$, the accumulated phase fluctuations become dominant, resulting in an effective standard quantum limit scaling,
\begin{equation}
\chi^{-2}\propto  \frac{N_e \langle \cos\phi \rangle^2}{\Delta^2 \text{Re}(g_se^{i\phi})} .
\end{equation}
The crossover is caused by the classical uncertainty of the electron ensemble rather than the quantum noise of the optical probe. 

\begin{figure}
    \centering
    \hspace{-1.0cm}
    \includegraphics[width=1.1\linewidth]{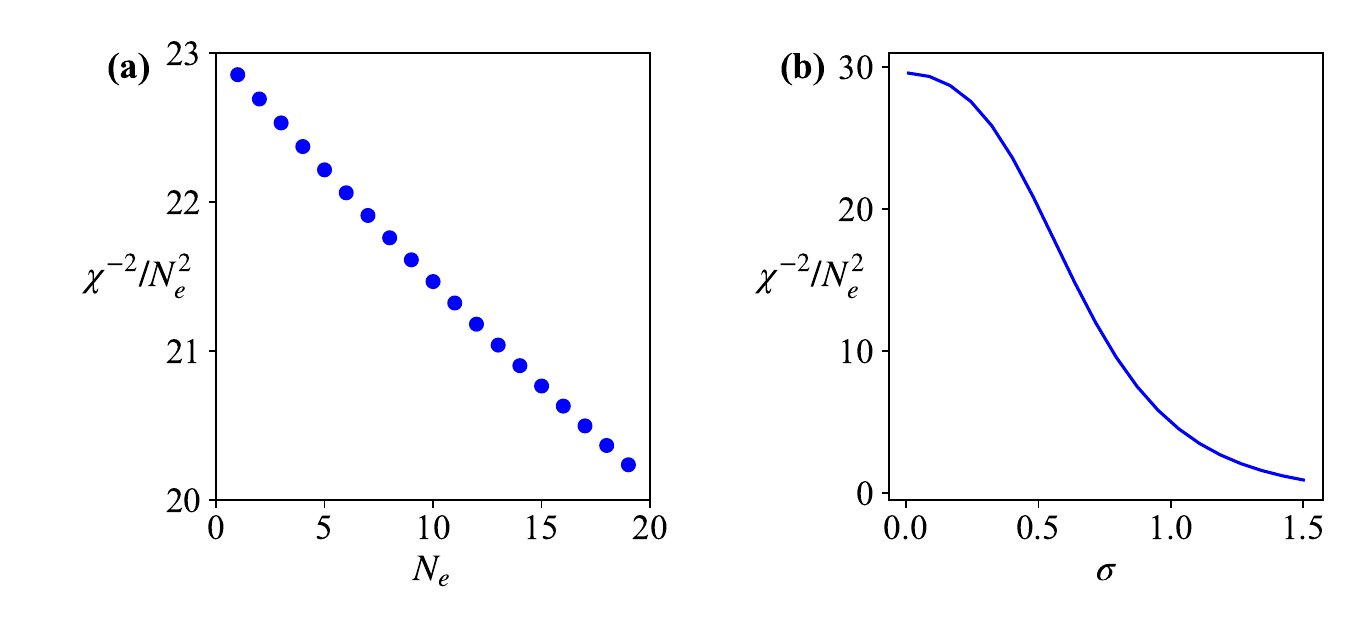}
   \caption{Degradation of the normalized homodyne sensitivity $\chi^{-2}/N_e^2$ induced by imperfect electron-state preparation, with the squeezing parameter of the optical mode fixed at $r=1.0$. (a) Dependence on the electron number $N_e$ at a fixed phase-distribution width $\sigma=0.5$. (b) Dependence on $\sigma$ at a fixed electron number $N_e=20$.  The accumulated phase fluctuations introduce a classical noise contribution that grows with $N_e$ and with $\sigma$, destroying the ideal $\chi^{-2}\propto N_e^{2}$ scaling and reducing the metrological advantage.}
    \label{fig:imperfect-electron}
\end{figure}

Figure~\ref{fig:imperfect-electron} illustrates the degradation of the sensitivity scaling induced by imperfect electron preparation. We consider a Gaussian phase distribution $p_\phi(\phi)\propto\exp[-\phi^2/(2\sigma^2)]$ with $\phi\in[-\pi,\pi)$, where $\sigma$ characterizes the phase uncertainty of the electron comb states. It is shown that phase fluctuations introduce an additional classical noise contribution that increases with the electron number, thereby destroying the ideal $N_e^2$ scaling and reducing the metrological advantage.

\begin{figure}
    \centering
    \includegraphics[width=1.0\linewidth]{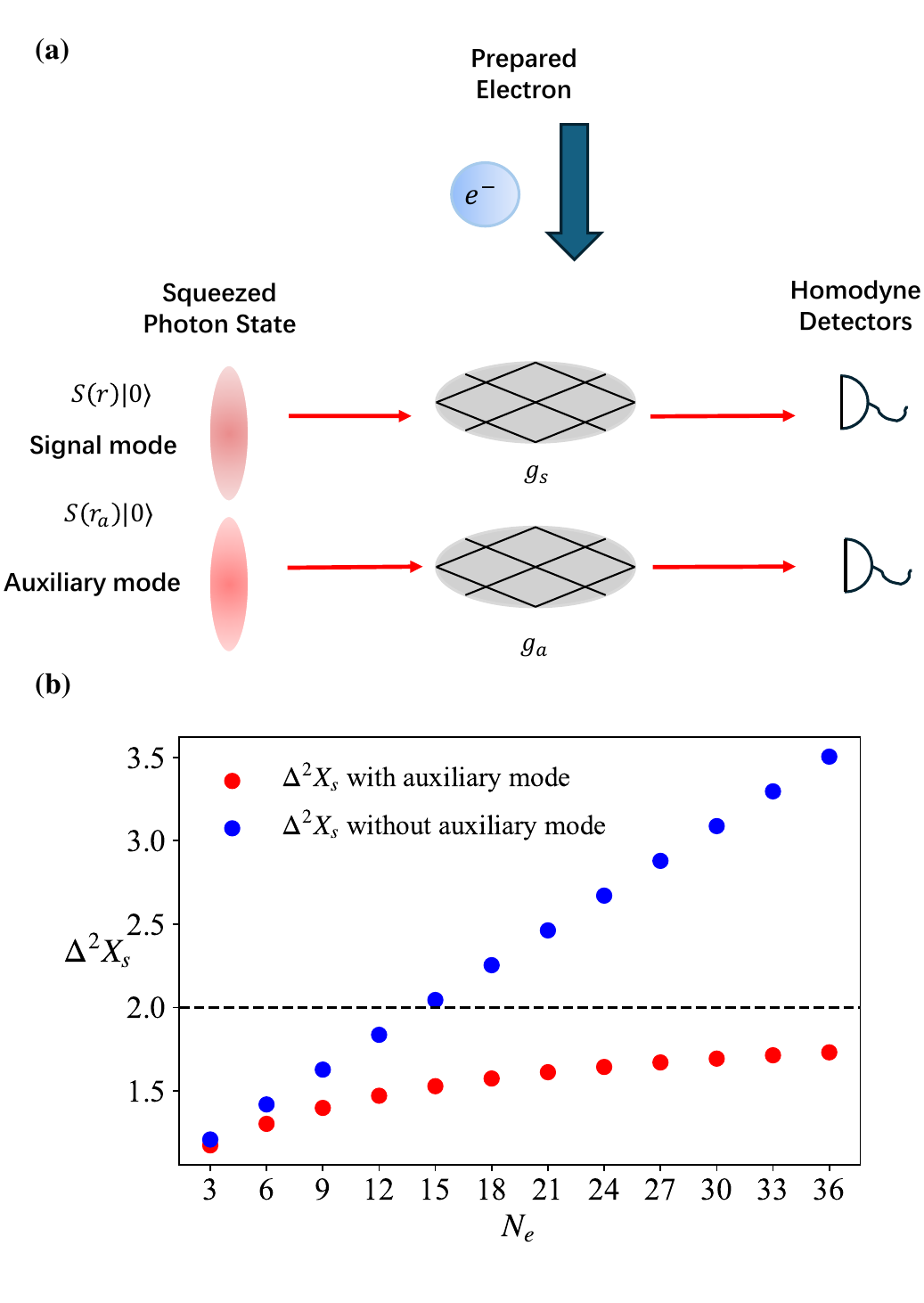}
   
    \caption{Recovery of Heisenberg scaling using an auxiliary optical mode. (a) The experimental apparatus of the recovery scheme. For the case of estimating $\text{Re}(g_s)$, the prepared electrons interact with the signal and the auxiliary modes with initial state $S(r)\ket 0$ and $S(r_a)\ket 0$. Then homodyne detection is performed to the two modes respectively. (b) The comparison of $\Delta^2 X_s$ between the schemes with and without the auxiliary mode, where $\Delta^2\cos\phi =0.436$, $g_s=g_a=0.2$ and $r=r_a=0$. For the scheme without the auxiliary mode, the $\Delta^2 X_s $ increases linearly with the electron number. However, for the scheme with auxiliary mode, the $\Delta^2 X_s $ has an upper bound related to the squeezing parameter of the signal and the auxiliary mode. }
    \label{fig:auxiliary}
\end{figure}

To overcome this limitation, we introduce an auxiliary optical mode that interacts with the same electron ensemble through an additional PINEM coupling of strength $g_a$. 
Here, to focus on the noise-suppression mechanism and the recovery of Heisenberg scaling without unnecessary technical complexity, we consider, without loss of generality, the case of $\text{Im}(g_s)=\text{Im}(g_a)=0$, in which only the amplitude-quadrature $X$ is measured to estimate $\text{Re}(g_s)$. 
Thus, as shown in Fig.~\ref{fig:auxiliary}(a), both the signal and the auxiliary modes are prepared in squeezed vacuum states $S(r)\ket 0$ and $S(r_a)\ket 0$, and interact with the imperfectly prepared electrons. In the following, we set $r=r_a$. After tracing out the electrons, the output state of the optical modes can be expressed as a correlated mixture,
\begin{equation}
\rho_{\rm ph,eff}
=
\sum_\eta p(\eta)
\rho_{\rm ph}(g_s\eta/2)
\otimes
\rho_{\rm ph}^{(a)}(g_a\eta/2),
\end{equation}
where $\eta=2\sum_{i=1}^{N_e}\cos\phi_i$, whose probability distribution $p(\eta)$ is inherited from the electron phase distribution $p_\phi$, and $\rho_{ph}(\xi)=D(\xi)S(r)\ket 0\bra 0S(r)^\dagger D(\xi)^\dagger $.  For large $N_e$, the probability of $\eta$ can be approximated by a Gaussian distribution according to the central limit theorem,
\begin{equation}
p(\eta) = \frac{1}{\sqrt{2\pi \sigma_\eta^2}} \exp\!\Big[-\frac{(\eta - \mu_\eta)^2}{2 \sigma_\eta^2} \Big],
\end{equation}
with $\mu_\eta = 2 N_e \langle \cos \phi \rangle$, and $\sigma^2_\eta = 4 N_e \Delta^2(\cos \phi)$, where $\langle \cdot \rangle$ and $\Delta^2$ are evaluated with respect to $p_\phi$.

By performing homodyne detection on the auxiliary mode to measure its amplitude quadrature and applying the Bayesian framework, the conditional probability distribution of $\eta$ with a given outcome $x_a$ can be obtained, where the resulting posterior variance is (See SM~\cite{supplementary} for details)
\begin{equation}
\sigma_{\rm post}^2
=
\left(
\frac{1}{\sigma_\eta^2}
+
\frac{g_a^2}{e^{-2r}}
\right)^{-1}.
\end{equation}
Consequently, the quadrature variance of the signal mode becomes $\Delta^2X_s
=
e^{-2r}
+
g_s^2\sigma_{\rm post}^2 $. We compare the values of $\Delta^2 X_s$ in the schemes with and without the auxiliary mode in Fig.~\ref{fig:auxiliary}(b), showing that, unlike the linear scaling of $\Delta^2 X_s\propto N_e$ without the auxiliary mode, the variance is suppressed to a constant level of $\Delta^2 X_s<e^{-2r}(1+g_s^2 /g_a^2)$ by introducing the auxiliary mode (see SM~\cite{supplementary} for details).
Recalling that the sensitivity in Eq.~(\ref{eq:imperfect_sensitivity}) can be rewritten as $\chi^{-2}=4N_e^2\langle\cos\phi\rangle^2/\Delta^2 X_s$, the Heisenberg scaling $\chi^{-2}\propto N_e^2$ can therefore be restored. 
This can be understood because the auxiliary optical mode serves as a syndrome channel that diagnoses the collective phase fluctuations arising from imperfect electron preparation. Since it interacts with the same electron ensemble as the signal mode, its homodyne result reveals the fluctuation-induced displacement of the signal mode. Conditioning the signal measurement on this information suppresses the associated classical noise and recovers the Heisenberg scaling of multi-electron PINEM metrology.
In conclusion, we establish a quantum metrological framework for estimating electron-photon coupling strengths in multi-electron PINEM. Electron coherence on the energy-comb basis converts the interaction into an optical displacement enhanced by the electron number, yielding the Heisenberg scaling in the electron number, $F_Q\propto N_e^2$. With squeezed optical modes and homodyne detection, this Heisenberg scaling can be achieved with an additional linear sensitivity in the photon number, exceeding conventional PINEM measurements even in single-electron cases. However, imperfect comb-state preparation induces electron phase fluctuations and associated classical noise that degrade the ideal scaling. To fill this gap, we show that an auxiliary optical mode, coupled to the same electron ensemble and subjected to conditional measurements, suppresses these fluctuations and thereby recovers the Heisenberg scaling for imperfect electron states. Our results identify free-electron coherence as a resource for quantum-enhanced sensing and provide a general strategy for mitigating state-preparation imperfections in electron-photon and broader quantum metrological platforms.

\begin{acknowledgments}
This work is supported by the Quantum Science and Technology-National Science and Technology Major Project (Grant Nos. 2025ZD0301000 and 2024ZD0302401), the National Natural Science Foundation of China (Grant Nos. 12125402, 12534016, 12334013, 92250306, and 12474256), Beijing Natural Science Foundation (Grant No. Z240007), and the State Key R\&D Program (Grant No. 2022YFA1604301). F.-X. S. acknowledges support from the Fundamental Research Funds for the Central Universities (Grant No. 539926010).
\end{acknowledgments}

\bibliography{ref}

\end{document}


\title{Supplemental Material:
Heisenberg-Scaling Quantum Metrology of Electron-Photon Coupling via Multi-Electron PINEM}

\author{Zihang Zou}
\affiliation{State Key Laboratory of Artificial Microstructure and Mesoscopic Physics, School of Physics, Frontiers Science Center for Nano-optoelectronics, $\&$ Collaborative Innovation Center of Quantum Matter, Peking University, Beijing 100871, China}
\author{Feng-Xiao Sun}
\email{sunfengxiao@bupt.edu.cn}
\address{State Key Laboratory of Information Photonics and Optical Communications, Beijing Key Laboratory of Information Metamaterials, $\&$ School of Physical Science and Technology, Beijing University of Posts and Telecommunications, Beijing 100876, China}
\address{State Key Laboratory of Artificial Microstructure and Mesoscopic Physics, School of Physics, Frontiers Science Center for Nano-optoelectronics, $\&$ Collaborative Innovation Center of Quantum Matter, Peking University, Beijing 100871, China}
\author{Yunquan Liu}
\address{State Key Laboratory of Artificial Microstructure and Mesoscopic Physics, School of Physics, Frontiers Science Center for Nano-optoelectronics, $\&$ Collaborative Innovation Center of Quantum Matter, Peking University, Beijing 100871, China}
\address{Collaborative Innovation Center of Extreme Optics, Shanxi University, Taiyuan, Shanxi 030006, China}
\author{Qiongyi He}
\email{qiongyihe@pku.edu.cn}
\address{State Key Laboratory of Artificial Microstructure and Mesoscopic Physics, School of Physics, Frontiers Science Center for Nano-optoelectronics, $\&$ Collaborative Innovation Center of Quantum Matter, Peking University, Beijing 100871, China}
\address{Collaborative Innovation Center of Extreme Optics, Shanxi University, Taiyuan, Shanxi 030006, China}
\address{Hefei National Laboratory, Hefei 230088, China}

\maketitle
\tableofcontents

\section{Quantum Fisher Information of Multi-Electron PINEM}
We first introduce the quantum Fisher information (QFI) to evaluate the metrological potential of our proposal. According to the quantum Cram\'{e}r-Rao bound~\cite{CR-bound}, the ultimate achievable sensitivity to estimate a parameter $d$ --- for a given probe state $\rho_0$ undergoing encoding $\rho_d=e^{id\hat{G}}\rho_0 e^{-id\hat{G}}$ with generator $\hat{G}$ --- is determined by the inverse of QFI,
\begin{equation}
\label{cr-b}
    \Delta^2 d \ge \Delta^2 d_{QCR} \equiv \frac{1}{mF_Q(\rho_0,\hat{G})},
\end{equation}
where $m$ is the number of independent repetitions which will be ignored in the following discussion, and the QFI corresponds to the maximal sensitivity over all possible measurements,
\begin{equation}
    F_Q(\rho_0,\hat{G}) = \sum_{jk}\frac{(\lambda_j-\lambda_k)^2}{\lambda_j+\lambda_k}|\langle j|\hat{G}|k\rangle|^2,
\end{equation}
with $\lambda_j$ and $\ket{j}$ being the eigenvalues and eigenstates of $\rho_0$, respectively. When the probe state is a pure state, the QFI can be simplified as $F_Q(\rho_0,\hat{G})=4\Delta^2 \hat{G}$, where $\Delta^2 \hat G = \langle \hat G^2 \rangle - \langle \hat G \rangle^2$. And for a specific measurement $M$, the sensitivity can be evaluated by the squeezing parameter~\cite{PhysRevA.46.R6797,PhysRevA.47.5138,MAI},
\begin{equation}
\chi^{2}\equiv \frac{\Delta^2 M}{|\partial_{d} \langle M\rangle |^2}\ge\frac{1}{F_Q}.
\end{equation}
$M$ is called the optimal measurement if the QFI is reached $\chi^{-2}=F_Q$. Hence, to enhance the accuracy of parameter estimation, a higher QFI is necessary.

We analyze the ultimate precision limit for estimating a complex electron-photon coupling $g_s = \operatorname{Re}(g_s) + i \operatorname{Im}(g_s)= g_x + i g_y$ in a multi-electron PINEM interaction. 
The scattering operator is
\begin{equation}
U(g_s) = \exp\Big[ \sum_{i=1}^{N_e} \big( g_s a^\dagger b_i - g_s^* a b_i^\dagger \big) \Big],
\end{equation}
where $a$ is the photon annihilation operator and $b_i$ is the energy-ladder operator of the $i$-th electron,
\begin{equation}
b_i |E\rangle_i = |E - \hbar\omega \rangle_i.
\end{equation}

For the purpose of parameter estimation, we treat the real and imaginary parts of \(g_s\) as independent parameters.
The generator associated with each parameter is obtained by taking derivatives of the unitary $U(g_s)$ with respect to $g_x$ or $g_y$,
\begin{align}
\hat G_x &= -i\frac{\partial}{\partial g_x} \left[ \sum_i (g_s a^\dagger b_i - g_s^* a b_i^\dagger) \right] 
=-i \sum_{i=1}^{N_e} (a^\dagger b_i - a b_i^\dagger), \\
\hat G_y &=-i \frac{\partial}{\partial g_y} \left[ \sum_i (g_s a^\dagger b_i - g_s^* a b_i^\dagger) \right] 
= \sum_{i=1}^{N_e} (a^\dagger b_i + a b_i^\dagger).
\end{align}
Hence, for a pure probe state \(|\psi\rangle\), the QFI for each parameter is given by
\begin{equation}
F_Q^{(x)} = 4 \, \Delta^2 \hat G_x, \qquad
F_Q^{(y)} = 4 \, \Delta^2 \hat G_y.
\end{equation}
For identical electrons prepared in the same quantum state, the QFI can be expanded as
\begin{align}
F_Q^{(x)} &= 4 N_e \Gamma_{ii}^{(x)} + 4 N_e (N_e - 1) \Gamma_{ij}^{(x)}, \\
F_Q^{(y)} &= 4 N_e \Gamma_{ii}^{(y)} + 4 N_e (N_e - 1) \Gamma_{ij}^{(y)},
\end{align}
where
\begin{align}
\Gamma_{ii}^{(x)} &= \langle (i(a^\dagger b_i - a b_i^\dagger))^2 \rangle - \langle i(a^\dagger b_i - a b_i^\dagger) \rangle^2, \\
\Gamma_{ij}^{(x)} &= \langle (i(a^\dagger b_i - a b_i^\dagger)) (i(a^\dagger b_j - a b_j^\dagger)) \rangle - \langle i(a^\dagger b_i - a b_i^\dagger) \rangle \langle i(a^\dagger b_j - a b_j^\dagger) \rangle.
\\
\Gamma_{ii}^{(y)} &= \langle (a^\dagger b_i + a b_i^\dagger)^2 \rangle - \langle a^\dagger b_i + a b_i^\dagger \rangle^2, \\
\Gamma_{ij}^{(y)} &= \langle (a^\dagger b_i + a b_i^\dagger)(a^\dagger b_j + a b_j^\dagger) \rangle - \langle a^\dagger b_i + a b_i^\dagger \rangle \langle a^\dagger b_j + a b_j^\dagger \rangle.
\end{align}
The inter-electron covariance terms $\Gamma_{ij}^{(x)}$ and $\Gamma_{ij}^{(y)}$ are responsible for the collective \(N_e^2\) scaling of the QFI. In this work we consider the electrons and the photon modes are initially fully separable then $\langle a^\dagger b_i\rangle=\langle a^\dagger\rangle \langle b_i\rangle$. Thus, the coherence of each electron in the ladder basis, quantified by \(\langle b_i \rangle \neq 0\), is a necessary condition for Heisenberg-like enhancement in both \(\operatorname{Re}(g_s)\) and \(\operatorname{Im}(g_s)\) estimation. In addition, noting that $\Gamma_{ij}^{(\mu)}$ ($\mu=x,y$) can reach the scaling of $N_{ph}=\langle a^\dagger a\rangle$, the quantum Fisher information $F_Q^{(\mu)}$ is thus in the scaling of $N_e^2 N_{ph}$. Therefore, the maximal scaling of $\chi^{-2}$ is $N_e^2 N_{ph}$, which can be attained by our proposed scheme in the main text.

\section{Analysis of Non-commutativity in the PINEM Operator }

It should be emphasized that the generators $G_x$ and $G_y$ do not commute. Thus, it becomes essential to analyze whether the ultimate quantum sensitivity can be obtained for both parameters $g_x$ and $g_y$ simultaneously. To fill this gap, we define $B=\sum_i b_i$ and then obtain
\begin{equation}
    [G_x,G_y]=2iB^\dagger B,
    \qquad
    [B^\dagger B,G_x]=[B^\dagger B,G_y]=0.
\end{equation}
The Baker-Campbell-Hausdorff identity therefore gives
\begin{equation}
    \exp[i(g_xG_x+g_yG_y)]
    =
    e^{ig_xg_yB^\dagger B}
    \exp(ig_xG_x)\exp(ig_yG_y).
\end{equation}
We explain in the following why the additional factor does not affect the optical output considered in our protocols.

When all electrons are prepared in comb states
$|\mathrm{comb}(\phi_i)\rangle$, the operator $B^\dagger B$
has the definite value $|\sum_i e^{i\phi_i}|^2$ on the
prepared electron state. The additional factor thus becomes
a phase of $e^{i\chi}$ with
\begin{equation}
    \chi=g_xg_y\left|\sum_i e^{i\phi_i}\right|^2.
\end{equation}
This overall phase does not affect
the output density matrix and can therefore be ignored.

For other electron state preparations, noting that the electrons are traced out without further operations in our protocols and only the resulting optical output is detected, the electron input can therefore be treated equivalently as a mixture of comb-state components, weighted by its diagonal elements in the comb-state basis. In this way, each component acquires its own scalar phase $e^{i\chi}$, which cancels in the corresponding optical density matrix,
\begin{equation}
    e^{i\chi}\rho_{\mathrm{ph}}e^{-i\chi}
    =
    \rho_{\mathrm{ph}}.
\end{equation}
Consequently, averaging over these components gives the same reduced optical state whether or not the additional phase factor is retained. The non-commutativity of $G_x$ and $G_y$ therefore does not introduce a correction to the optical output used in our measurement protocols. For similar reason, the effect of non-commutativity among the electrons can be ignored.

\section{Comb State Electrons Interact With Photon mode}

The physical origin of the quadratic enhancement can be illustrated using electron comb states, which take the form of
\begin{equation}
    |\mathrm{comb}(\phi)\rangle
    =
    \sum_{n=-\infty}^{+\infty}
    e^{in\phi}|n\rangle.
\end{equation}
Please note that the electron comb state is the eigenstate of the energy-ladder operator $b$, that is,
\begin{equation}
    b|\mathrm{comb}(\phi)\rangle
    =
    e^{i\phi}
    |\mathrm{comb}(\phi)\rangle .
\end{equation}
In particular, we assume that the electrons are prepared in identical comb states with the phases $\phi=0$, which satisfy
\begin{equation}
    b_j|\mathrm{comb}(0)\rangle_j
    =
    |\mathrm{comb}(0)\rangle_j.
\end{equation}
Thus, the initial state of the whole system can be expressed as
\begin{equation}
    |\Psi_0\rangle
    =
    |\mathrm{comb}(0)\rangle^{\otimes N_e}
    \otimes
    |\psi_{\rm ph}\rangle.
\end{equation}
Since each electron is prepared in the eigenstate of \(b_i\) with the eigenvalue \(1\), we have
\begin{equation}
    \sum_{i=1}^{N_e}
    \left(
    g_s a^\dagger b_i
    -
    g_s^* a b_i^\dagger
    \right)
    |\Psi_0\rangle
    =
    \left(
    N_e g_s a^\dagger
    -
    N_e g_s^* a
    \right)
    |\Psi_0\rangle .
\end{equation}
Therefore, the final state after the PINEM interaction can be obtained as
\begin{equation}
    |\Psi_f\rangle
    =U(g_s)
    |\Psi_0\rangle
    =
    |\mathrm{comb}(0)\rangle^{\otimes N_e}
    \otimes
    D(N_e g_s)
    |\psi_{\rm ph}\rangle ,
\end{equation}
where
\begin{equation}
    D(\alpha)
    =
    \exp(\alpha a^\dagger-\alpha^* a)
\end{equation}
is the displacement operator for the optical mode. Thus, for electrons prepared in identical zero-phase comb states, the multi-electron PINEM interaction is equivalent to an optical displacement with an amplitude collectively enhanced by \(N_e\). In particular, the real and imaginary components of the coupling, \(g_x=\mathrm{Re}(g_s)\) and \(g_y=\mathrm{Im}(g_s)\), are encoded in the displacement \(N_e g_s\). 

The corresponding generators are proportional to \(N_e i(a^\dagger-a)\) and \(N_e(a+a^\dagger)\) for estimating \(g_x\) and \(g_y\), respectively. Consequently, the QFIs satisfy
\begin{equation}
    F_Q^{(x)}
    \propto
    N_e^2,
    \qquad
    F_Q^{(y)}
    \propto
    N_e^2.
\end{equation}
This result directly establishes Heisenberg scaling with the number of electrons in multi-electron PINEM metrology.

\section{Metrology Protocol in the Ideal Regime}

In our scheme, the initial state of light is set as a squeezed vacuum state, $|\psi^{(\pm)}_{\rm ph}\rangle_0=S(\pm r) |0\rangle$. Thus, the outgoing optical state becomes
\begin{equation}
|\psi^{(\pm)}_{\rm ph}\rangle_f=D(N_e g_s) S(\pm r) |0\rangle.
\end{equation}

Then the real and imaginary parts of $g_s$ are estimated independently through homodyne detection on the outgoing optical states. For the squeezed state $|\psi^{(+)}_{\rm ph}\rangle_f=D(N_e g_s) S(r)|0\rangle$, we measure the amplitude-quadrature
\begin{equation}
X = a + a^\dagger,
\end{equation}
whose expectation value is proportional to the real part of $g_s$,
\begin{equation}
\langle X \rangle = 2 N_e \operatorname{Re}(g_s)= 2 N_e g_x.
\end{equation}
In addition, the corresponding variance is determined by the squeezing degree of the optical state,
\begin{equation}
\Delta^2 X = e^{-2r}.
\end{equation}

Similarly, for $|\psi^{(-)}_{\rm ph}\rangle_f=D(N_e g_s) S(-r)|0\rangle$, we measure the phase-quadrature
\begin{equation}
P = i(a^\dagger - a),
\end{equation}
yielding
\begin{equation}
\langle P \rangle = 2 N_e \operatorname{Im}(g_s)= 2 N_e g_y, \quad \Delta^2 P=e^{-2r},
\end{equation}
which are associated with the imaginary part of $g_s$ and the squeezing parameter, respectively.

The corresponding sensitivity can be directly obtained,
\begin{align}
\chi^{-2}_{x} &= \frac{ |\partial_{g_x} \langle X \rangle |^2 }{ \Delta^2 X } = 4 e^{2r} N_e^2\propto N_e^2 N_{ph},\label{eq:sensitivity_comb}\\
\chi^{-2}_{y} &= \frac{ |\partial_{g_y} \langle P \rangle |^2 }{ \Delta^2 P } = 4 e^{2r} N_e^2\propto N_e^2 N_{ph}.
\end{align}
Obviously, $\chi^{-2}=\chi^{-2}_{x}=\chi^{-2}_{y}$ saturates the quantum Cram\'{e}r-Rao bound.

    \section{The Variance of Homodyne Detection under Imperfect Electron-State Preparation}

We start with an arbitrary initial state of electrons, where each electron is prepared in the state $\rho_e$ and the initial photon state is $\rho_{ph}$, the final electron-photon state after the PINEM interaction is 
\begin{equation}
    \rho=U(g_s)[\rho_e^{\otimes N_e}\otimes \rho_{ph}]U(g_s)^\dagger,
\end{equation}
with
\begin{equation}
    U(g_s) = \exp\Big[ \sum_{i=1}^{N_e} \big( g_s a^\dagger b_i - g_s^* a b_i^\dagger \big) \Big].
\end{equation}
By tracing out the electron state, the final photon state is obtained,
\begin{equation}
    \rho_{ph}^{(f)}=\sum_{\vec \phi} \bra{\text{comb}(\vec\phi)} U(g_s)[\rho_e^{\otimes N_e}\otimes \rho_{ph}]U(g_s)^\dagger\ket {\text{comb}(\vec\phi)},
\end{equation}
where $\vec \phi=(\phi_1,\phi_2,...\phi_{N_e})$ denotes the phase of $N_e$ electrons and $\ket{\text{comb}(\vec\phi)}=\bigotimes_{i}\ket{\text{comb}(\phi_i)}.$ Considering that 
\begin{equation}
    U(g_s) \ket {\text{comb}(\vec\phi)}=D\left(g_s\sum e^{i\phi_i} \right) \ket {\text{comb}(\vec\phi)},
\end{equation}
the final photon state can be reduced to

\begin{equation}
      \rho_{ph}^{(f)}=\sum_{\vec \phi} \bra{\text{comb}(\vec\phi)}\rho_e^{\otimes N_e} \ket{\text{comb}(\vec\phi)} \otimes D(g_s\sum e^{i\phi_i} ) \rho_{ph} D(g_s\sum e^{i\phi_i} )^\dagger.
\end{equation}
Note that $\bra{\text{comb}(\vec\phi)}\rho_e^{\otimes N_e} \ket{\text{comb}(\vec\phi)} =\prod_i \bra{\text{comb}(\phi_i)}\rho_e \ket{\text{comb}(\phi_i)}=p(\vec \phi)  $, which defines the joint probability distribution of the phase vector $\vec\phi$. Hence, the relevant properties
of the electron state $\rho_e$ are fully specified by its diagonal
elements in the comb-state basis, $p_\phi= \bra{\text{comb}(\phi)}\rho_e \ket{\text{comb}(\phi)}$, as introduced in the main text.

Here, we take the estimation of $g_x=\text{Re}(g_s)$ as an example. As shown in the main text, the initial optical state in our scheme is the squeezed vacuum state $\rho_{ph}= S(r)\ket 0\bra 0 S(r)^\dagger$. In order to obtain the variance $\Delta^2X$ of $\rho_{ph}^{(f)}$, we firstly calculate the moments, 
\begin{align}
    \langle X\rangle&=\text{Tr}(  \rho_{ph}^{(f)} X)=\sum_{\vec \phi} p({\vec \phi})\text{Tr}[ (\rho_{ph} D(g_s\sum e^{i\phi_i} )^\dagger) XD(g_s\sum e^{i\phi_i} )]
    \\
    &=\sum_{\vec \phi}2p(\vec \phi)\text{Re}(g_s\sum_i e^{i\phi_i}),
    \\
    \langle X^2\rangle&=\text{Tr}(  \rho_{ph}^{(f)} X^2)=\sum_{\vec \phi} p({\vec \phi})\text{Tr}[ (\rho_{ph} D(g_s\sum e^{i\phi_i} )^\dagger) X^2D(g_s\sum e^{i\phi_i} )]
    \\
    &=\sum_{\vec \phi}4p(\vec \phi)\text{Re}(g_s\sum_i e^{i\phi_i})^2+ e^{-2r}.
\end{align}
It follows that $\Delta^2 X= e^{-2r} + 4\Delta^2 \text{Re}(g_s\sum_i e^{i\phi_i})$. For independently and identically prepared electrons, the variance is reduced to $\Delta^2 X= e^{-2r} + 4N_e \Delta^2 \text{Re}(g_se^{i\phi})$, where $\Delta^2 \text{Re}(g_se^{i\phi})  $ is the variance of $\text{Re}(g_se^{i\phi})$ with respect to the probability distribution $p_\phi$. And the $|\partial \langle X\rangle/\partial g_s|$ can be reduced to $2\sum_{\vec \phi} p(\vec \phi)  \sum_i \cos\phi_i =2N_e\langle \cos\phi\rangle$. Thus, the resulting sensitivity for estimating $\text{Re}(g_s)$ is
\begin{equation}
\chi^{-2}
=
\frac{
4N_e^2\langle\cos\phi\rangle^2
}{
e^{-2r} + 4N_e \Delta^2 \text{Re}(g_se^{i\phi})
}.
\label{eq:imperfect_sensitivity}
\end{equation}

\section{Recovery of Heisenberg Scaling Using an Auxiliary Optical Mode}

\subsection*{Joint Electron-Photon State}

By introducing an auxiliary optical mode that interacts with the same electron ensemble through an additional PINEM coupling $g_a$, the joint electron-photon density matrix after the interaction becomes
\begin{align}
\rho_{\rm ele-ph} =& \sum_{\vec \phi} p(\vec \phi) \, 
\left(
\bigotimes_{i=1}^{N_e} |\mathrm{comb}(\phi_i)\rangle \langle \mathrm{comb}(\phi_i)|
\right)\otimes
\rho_{\rm ph}\Big(g_s \sum_i e^{i\phi_i}\Big)
\otimes
\rho_{\rm ph}^{(a)}\Big(g_a \sum_i e^{i\phi_i}\Big),
\end{align}
where $\vec \phi = (\phi_1, \dots, \phi_{N_e})$. Since the electrons are not measured, we trace over the electronic degrees of freedom to obtain the effective optical state,
\begin{equation}
\rho_{\rm ph,eff} = \sum_{\vec \phi} p(\vec \phi) \,
\rho_{\rm ph}\Big(g_s \sum_i e^{i\phi_i}\Big)
\otimes
\rho_{\rm ph}^{(a)}\Big(g_a \sum_i e^{i\phi_i}\Big).
\end{equation}
In the following discussion, we assume both PINEM coupling strengths $g_s$ and $g_a$ to be real numbers for simplicity. 

\subsection*{Conditional Measurement and Posterior Variable}

We define a new random variable $\eta$ to describe the collective effect of measurement in the auxiliary mode,
\begin{equation}
\eta = 2 \sum_{i=1}^{N_e} \cos \phi_i,
\end{equation}
which is distinct from the random variable $\xi$ used in the main mode analysis. 
Each $\phi_i$ is independently sampled from $p_\phi$, so $\eta$ is a random variable under the product distribution $p(\vec \phi) = \prod_i p_\phi(\phi_i)$. 
Since both optical modes depend on the electron ensemble only through the collective variable $\eta$, the effective two-mode optical state can be rewritten as
\begin{equation}
\rho_{\rm ph,eff}
=
\sum_{\eta}
p(\eta)\,
\rho(\eta),
\end{equation}
where 
\begin{equation}
\rho(\eta)
=
\rho_{\rm ph}(\eta g_s/2)
\otimes
\rho_{\rm ph}^{(a)}(\eta g_a/2),
\end{equation}
and $p(\eta)$ denotes the probability distribution induced by the electron phase distribution $p_\phi$.

In the large $N_e$ limit, the prior distribution of $\eta$ approaches a Gaussian distribution according to the central limit theorem,
\begin{equation}
p(\eta) = \frac{1}{\sqrt{2\pi \sigma_\eta^2}} \exp\!\Big[-\frac{(\eta - \mu)^2}{2 \sigma_\eta^2} \Big],
\end{equation}
with
\begin{equation}
\mu_\eta = 2 N_e \langle \cos \phi \rangle,
\qquad
\sigma^2_\eta = 4 N_e \Delta^2(\cos \phi),
\end{equation}
where $\langle \cdot \rangle$ and $\Delta^2$ are defined according to the prior $p_\phi$.

Performing homodyne detection on the auxiliary mode corresponds to measuring the quadrature operator
\[
X_a=a_a+a_a^\dagger
\]
on each conditional state $\rho(\eta)$. For the auxiliary mode prepared in state $S(r_a)\ket 0$, the resulting measurement probability is
\begin{equation}
P(x|\rho(\eta))
\propto
\exp\!\left[
-\frac{(x-g_a\eta)^2}
{2e^{-2r_a}}
\right],
\end{equation}
where $e^{-2 r_a}$ is the quadrature variance of the auxiliary squeezed mode. Therefore, if the measured result is $X_a=x$, the conditional state of the main optical mode is 
\begin{equation}
    \rho_s(x)=\sum_\eta P(\eta|x)\rho_{ph}(g_s\eta/2),
\end{equation}
where the posterior distribution of $\eta$ is then
\begin{equation}
P(\eta|x) \propto P(x|\rho(\eta))\,p(\eta).
\end{equation}
The distribution $P(\eta|x) $ remains Gaussian with the posterior mean,
\begin{equation}
\mu_{\rm post}
=
\frac{
\mu_\eta+
g_a\sigma_\eta^2 e^{2r_a}x
}{
1+g_a^2\sigma_\eta^2 e^{2r_a}
}
\end{equation}
and the posterior variance,
\begin{equation}
\sigma_{\rm post}^2 = \left( \frac{1}{\sigma_\eta^2} + \frac{g_a^2}{e^{-2 r_a}} \right)^{-1} 
= \frac{\sigma_\eta^2 e^{-2 r_a}}{\sigma_\eta^2 g_a^2 + e^{-2 r_a}}<\frac{e^{-2r_a}}{g_a^2}.
\end{equation}
Consequently, the quadrature variance of the signal mode becomes
\begin{equation}
\Delta^2 X_s = e^{-2 r} + g_s^2 \sigma_{\rm post}^2,
\end{equation}
which is of constant scaling of $N_e$. In the main text, we consider the case $r=r_a$ which means the signal and auxiliary mode have the same squeezing parameter so that $\Delta^2X_s<e^{-2r}(1+g_s^2/g_a^2)$ . In this case, when $X_s=x_s$ and $X_a=x$ are measured, the coupling strength $g_s$ is estimated as,
\begin{equation}
    g_s=\frac{x_s}{\mu_{\rm post}}=\frac{x_s(1+g_a^2\sigma_\eta^2 e^{2r_a})}{\mu_\eta+
g_a\sigma_\eta^2 e^{2r_a}x},
\end{equation}
where the homodyne sensitivity is restored to Heisenberg scaling,
\begin{equation}
\chi^{-2} = \frac{\left| \frac{d \langle X_s \rangle}{d g_s} \right|^2}{\Delta^2 X_s}= \dfrac{4N_e^2\langle\cos\phi\rangle^2}{e^{-2r}+g_s^2\sigma^2_{\rm post}}\propto N_e^2.
\end{equation}

\section{Single-Electron Comparison with Conventional PINEM Schemes}

In the conventional scheme, the electron is initially prepared in an energy eigenstate and interacts with a coherent optical field of amplitude $\alpha$. Hence, the output electron-photon state after the PINEM interaction is
\begin{equation}
|\Psi\rangle = U(g_s) \, |\alpha\rangle_{\rm ph} \otimes |0\rangle_e,
\end{equation}
and the probability of detecting the electron in the $n$-th energy-ladder state is
\begin{equation}
P_n = |J_n(2 |g_s\alpha|)|^2,
\end{equation}
where $J_n$ is the Bessel function of the first kind. Without loss of generality, we assume that $g$ and $\alpha$ are real.

The inverse squeezing parameter (or sensitivity) can then be defined as
\begin{equation}
\chi^{-2} = \frac{|\partial_{g_s} P_n|^2}{\Delta^2 P_n}, \quad \Delta^2 P_n = P_n(1-P_n).
\end{equation}
Using the large-argument asymptotic expansion of the Bessel function,
\begin{equation}
J_n(x) \approx \sqrt{\frac{2}{\pi x}} \cos\Big(x - \frac{n\pi}{2} - \frac{\pi}{4}\Big), \quad x \gg 1,
\end{equation}
the fast oscillating factor can be separated from the slowly varying envelope. Then the sensitivity becomes
\begin{equation}
\chi^{-2} \approx \frac{16 \alpha}{\pi g_s} \sin^2\Big(2 g_s \alpha - \frac{n \pi}{2} - \frac{\pi}{4}\Big),
\end{equation}
where the oscillatory term $\sin^2(\cdot)$ varies rapidly with $n$ and $g_s$, while the prefactor $(16 \alpha)/(\pi g_s)$ defines the envelope. As a result, the envelope grows as
\begin{equation}
\mathrm{Env}(\chi^{-2}) \sim \alpha \sim \sqrt{N_{\rm ph}},
\end{equation}
demonstrating that even for large photon numbers, the sensitivity of the conventional scheme scales only as $\chi^{-2} \sim \sqrt{N_{\rm ph}}$.

By contrast, a single comb-state electron interacting with a squeezed optical mode already yields a sensitivity that scales linearly with the photon number,
\begin{equation}
\chi^{-2}_{\rm comb} \propto N_{\rm ph}.
\end{equation}
This can be obtained directly from Eq.~(\ref{eq:sensitivity_comb}).

\bibliography{ref}